# Direct observation of six-fold exotic fermions in topological semimetal PdSb$_2$

Z. P. Sun,[1,2,*] C. Q. Hua,[3,4,*] X. L. Liu,[5,*] Z. T. Liu,[1] M. Ye,[1] S. Qiao,[1]
Z. H. Liu,[1] J. S. Liu,[1] Y. F. Guo,[5,†] Y. H. Lu,[3,4,‡] and D. W. Shen[1,2,§]

[1]*Center for Excellence in Superconducting Electronics,
State Key Laboratory of Functional Materials for Informatics,
Shanghai Institute of Microsystem and Information Technology,
Chinese Academy of Sciences, Shanghai 200050, China*
[2]*Center of Materials Science and Optoelectronics Engineering,
University of Chinese Academy of Sciences, Beijing 100049, China*
[3]*Zhejiang Province Key Laboratory of Quantum Technology and Device,
Department of Physics, Zhejiang University, Hangzhou 310027, P. R. China*
[4]*State Key Lab of Silicon Materials, School of Materials Science and Engineering, Zhejiang University, Hangzhou 310027, P. R. China*
[5]*School of Physical Science and Technology, ShanghaiTech University, Shanghai 200031, China*

Pyrite-type PdSb$_2$ with a nonsymmorphic cubic structure has been predicted to host six-fold-degenerate exotic fermions beyond the Dirac and Weyl fermions. Though magnetotransport measurements on PdSb$_2$ suggest its topologically nontrivial character, direct spectroscpic study of its band structure remains absent. Here, by utilizing high-resolution angle-resolved photoemission spectroscopy, we present a systematic study on its bulk and surface electronic structure. Through careful comparison with first-principles calculations, we verify the existence of six-fold fermions in PdSb$_2$, which are formed by three doubly degenerate bands centered at the $R$ point in the Brillouin zone. These bands exhibit parabolic dispersion close to six-fold fermion nodes, in sharp contrast to previously reported ones in chiral fermion materials. Furthermore, our data reveal no protected Fermi arcs in PdSb$_2$, which is compatible with its achiral structure. Our findings provide a remarkable platform for study of new topological fermions and indicate their potential applications.

In high-energy physics, three types of fermions have been predicated to exist in our universe according to the Lanrence invarience, i.e. the Dirac, Weyl and Majorana fermions. Although the existence of Weyl and Majorana fermions as elementary particles is still under hot debate, some quasiparticles constructed out of excitations were theoretically proposed to be able to mimic all three fermions in solids. Very soon after these predictions, Dirac [1–4], Weyl [5–10] and Majorana fermions [11–13] were experimentally confirmed to actually exist in topological non-trivial materials. Moreover, rather than the Poincarè symmetry, it is the invariance under crystal symmetry of one of 230 space groups that constrains fermions in solids [14–17], and thus there might exist some novel fermions beyond elementary particles discussed in high-energy physics. In solids, it is well known that the topological non-trivial four- and two-fold degenerate band crossings correspond to Dirac and Wely fermions, respectively. While, allowed by crystal symmetries, some new exotic fermions showing three-, six-, or eight-fold degeneracy of band crossings have been proposed, which are particularly intriguing since there are no counterparts in high-energy physics due to Poincarè symmetry constrains [14].

To date, angle-resolved photoemission spectroscopy (ARPES) experiments have verified several types of exotic fermions beyond the Dirac and Weyl ones. Unique triply degenerate points in the electronic structure of molybdenum phosphide were reported [18]. Moreover, four-, six- (double spin-1) and eight-fold (charge-2) degenerate chiral fermions were as well observed in AlPt [19] and CoSi [20–23] families with extremely long surface Fermi arcs, in sharp contrast to Weyl semimetals which have multiple pairs of Weyl nodes with only small separation. In addition, four types of symmetry-stabilized topological fermions have been observed in PdBiSe [24]. The pyrite PdSb$_2$ is a superconductor ($T_c \sim 1.25K$) which has long been suggested to host six-fold degenerate exotic fermions beyond Dirac and Weyl fermions [25]. Recently, de Haasvan Alphen oscillations imposed with a nearly zero effective electron mass and a nontrivial Berry phase were detected when the magnetic field is applied along the [111] direction of crystalline PdSb$_2$ [26], likely in good consistence with the previous prediction of topological non-tivial exotic fermions therein. However, direct spectroscopic evidence for the six-fold fermion in PdSb$_2$ has been yet absent.

In this Letter, we performed comprehensive ARPES measurements on PdSb$_2$ single crystals to investigate their surface and bulk band structures, aided with density functional theory (DFT) calculations to interpret the results. High-resolution ARPES measurements on samples with well cleaved (001) planes unambiguously reveal the six-fold crossing around the time-reversal invariant $R$ point, protected by its nonsymmorphic symmetry: Pa$\bar{3}$, harmonically consistent with the calculations, thus evidently confirming the existence of six-fold fermions in PdSb$_2$.

High quality PdSb$_2$ single crystals were grown via the self-flux method similar as that described in [27]. As-grown crystals were carefully examined by powder X-ray diffraction (XRD) measurements using a Bruker D8 VENTURE single crystal diffractometer (Mo K$\alpha$1 radiation; $\lambda = 0.7093$ Å) at room temperature. ARPES measurements were performed at 03U beamline of Shanghai Synchrotron Radiation Facility (SSRF), 13U beamline of National Synchrotron Radiation Laboratory (NSRL) and One-Squared ARPES endstation at BESSY II. Energy and angular resolutions were set to around 10 meV and 0.2°, respectively. All samples were cleaved under a vacuum better than $5 \times 10^{-11}$ Torr. During measurements, the temperature was kept at 24 K. DFT calculations were performed in the Vienna *ab initio* package [28] by implementing projector-augmented wave (PAW) [29] pseudopotential with exchange-correlation functional of generalized gradient approximation of Perdew, Burke, and Ernzerhof (PBE) [30]. The convergence criteria for energy and force are set to $10^{-6}$ eV and 0.01 eV/Å, respectively. We dopted $13\times13\times13$ $k$-mesh for the first Brillouin zone. Maximally localized Wannier functions [31] based on Sb-$p$ and Pd-$d$ orbitals were generated and then using Wanniertool [32] to botain the surface state spectrum.

---

[*] Equal contributions
[†] guoyf@shanghaitech.edu.cn
[‡] luyh@zju.edu.cn
[§] dwshen@mail.sim.ac.cn

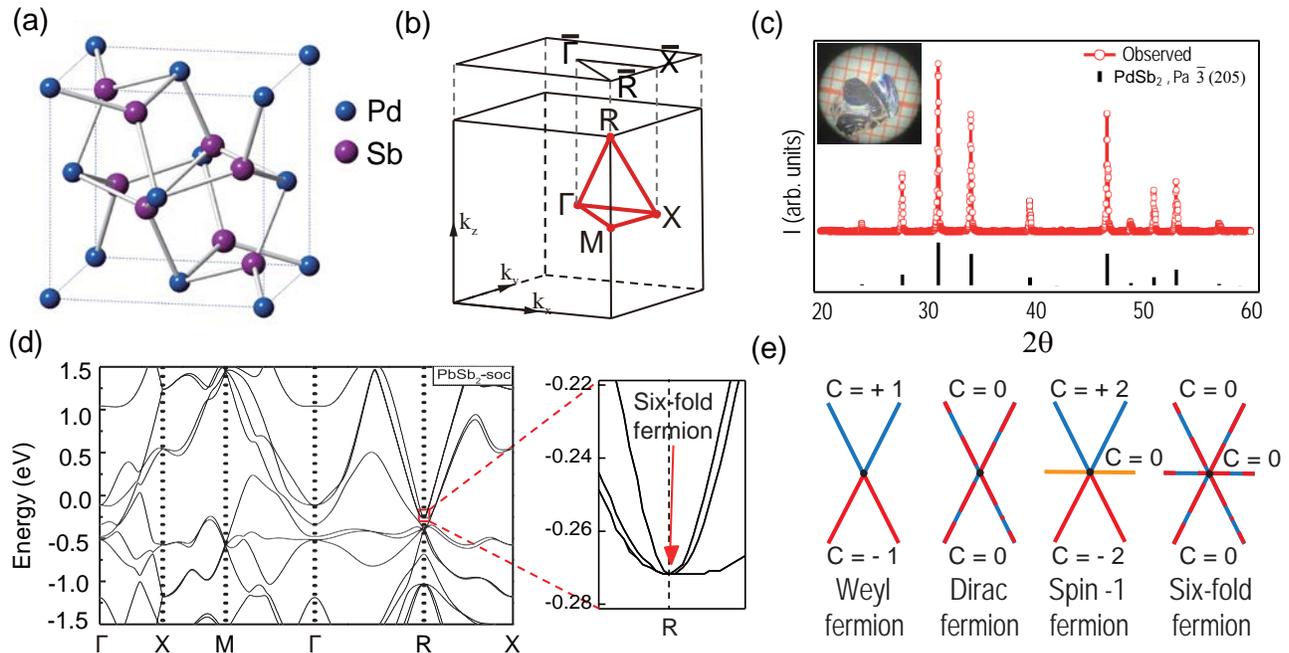

FIG. 1. Crystal structure and calculated band structure of cubic PdSb$_2$. (a) 3D crystal structure of cubic PdSb$_2$. (b) Bulk Brillouin zone and (001) surface Brillouin zone of PdSb$_2$ with high-symmetry momentum points marked. (c) Power x-ray diffraction (XRD) Pattern of PdSb$_2$ (observed pattern, red line; calculation with pyrite structure type [Pa3, space group 205], black line). The inset shows the photo of a typical PdSb$_2$ single crystal. (d) Band structure in the presence of SOC. The area marked by a red box is enlarged to reveal the six-fold degeneracy point. (e) Schematics of the band structures of a Weyl fermion, a Dirac fermion, a spin -1 fermion and a quadratic six-fold fermion.

The pyrite-type structure of cubic PdSb$_2$ is displayed in Fig. 1(a), in which blue and purple spheres represent Pd and Sb atoms, respectively. The corresponding three-dimensional Brillouin zones (BZ) and its projection along the (001) direction are shown in Fig. 1(b). Our single crystal has a typical size of $3 \times 2 \times 2$ $mm^3$, and it shows flat and shining surface after cleaving in the air, as illustrated by the inset of Fig. 1(c). Considering the sensitivity of electronic structure to the crystal structure and stoichiometry, we performed detailed powder X-ray diffraction (XRD) and energy dispersive X-ray spectrometer (EDS) measurements to characterize our samples. We found that the acquired powder XRD patterns match well with the calculated pyrite structure type of PdSb$_2$ with Pa3 (space group 205) as shown in Fig. 1(c). Moreover, we confirmed that the molar ratio of Pd:Sb = 1:2 of our samples used in this study (see more details in Supplemental Material [33]). Since this compound is not layer-structured, it could be cleaved along any high-symmetry directions. Thus, we performed X-ray Laue characterizations after ARPES investigation, and we could confirm that the cleavage planes are always the (001) planes (Fig. S1). Fig. 1(d) presents the bulk electronic structure calculation including the spin-orbit coupling (SOC). Consistent with previous prediction [14, 26], there exists one six-fold degeneracy at the $R$ point of the BZ, as highlighted by the zoom-in figure. Here, there are three bands congregating at the same point which is guaranteed by the $D_{3d}$ point group symmetry. Moreover, this three-fold degeneracy would be doubled by the presence of time-reversal (TR) symmetry [14]. Thus, in this compound, the six-fold degeneracy is expected to occur at the TR invariant $R$ point of the BZ. We note that such a six-fold fermion intrinsically different from double spin-1 chiral fermions [Fig. 1(e)] observed in CoSi families, which are guaranteed by the $T$ point group symmetry instead of the $D_{3d}$ symmetry [14, 21–23] and doubled by the presence of time-reversal (TR) symmetry. Unlike the spin-1 chiral fermions that carry non-zero Chern numbers of $\pm 2$, non-Abelian Berry curvature is expected in PdSb$_2$ [14].

Since the predicted six-fold degeneracy was suggested to be located at the $R$ point and the natural cleavage surface of our PdSb$_2$ sample corresponds to the (001) plane, as shown in Fig. 1(b), a detailed photon-energy-dependent ARPES measurement on $k_x$-$k_y$ plane is highly desired to identify the precise $k_z$ momentum position. Figs. 2(a) and 2(b) show photoemission intensity maps in the the $k_x$-$k_z$ plane taken at $E_F$ and $E_F - 0.1 eV$, respectively. One can clearly distinguish the pronounced periodic modulation of spectral feature along the $k_z$ direction. For the sake of comparison, the calculated band dispersion in the vicinity of $E_F$ along the same high symmetry direction is shown in Fig. 2(c), in which only six-fold degenerate electron-like bands around $R$ periodically intersect the Fermi level along $R - M - R$. Thus, we can establish the one-to-one relationship between our photoemission spectra and high-symmetry points along $k_z$. According to the free-electron final-state model [34],

$$k_z = \sqrt{\frac{2m}{\hbar^2}[(h\upsilon - \phi - E_b)cos^2\theta + V_0]}$$

where $h\upsilon$, $\varphi$, $V_0$ and $\theta$ represent the photon energy, work function, inner potential and emission angle of photoelectrons relative to the sample's surface normal, respectively, an inner potential of 28.0 eV could be obtained through the best fit to the periodic variation, with the lattice constant c = 6.554 Å and $\varphi$ = 4.25 eV. This result further allows us to determine that $h\upsilon$ = 20 and 48 eV are close to the $R$ points while 33 eV photons corresponds to the $M$ point. We note that our $k_z$ determination could be further confirmed by the reasonable consistency between experimental and corresponding constant energy contour calculation taken at different binding energies [black solid lines in Figs. 2(a) and 2(b)].

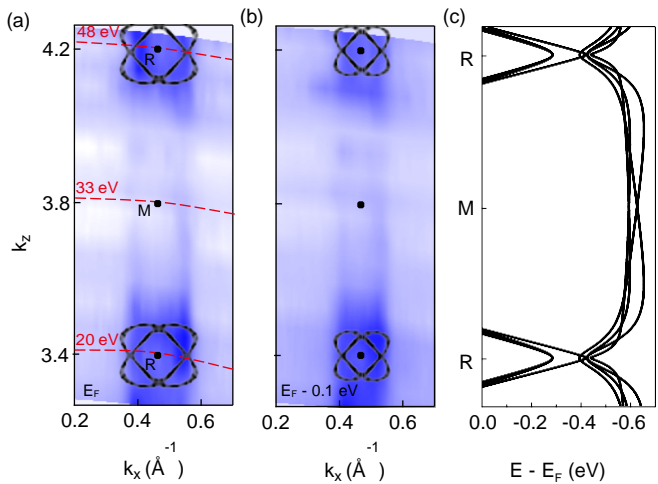

FIG. 2. The photon energy dependent ARPES data. (a), (b) Integrated photoemission intensity plots of $k_z$-$k_x$ plane, taken at $E_F$ and $E_F$ - 0.1 eV, respectively. The photon energies used for the $k_z$ dispersion measurement are from 16 to 50 $eV$. Periodically symmetry points $M$ and $R$ are marked corresponding specific energies, and the corresponding calculated bulk Fermi surfaces on the $k_z$-$k_x$ plane are also attached for the convenience of comparison. (c) Calculated band dispersion along $M - R - M$ high symmetry direction.

Considering the more bulk sensitivity of high-energy phonons in photoemission probing, we applied a linearly polarized $h\upsilon$ = 83 eV phonon to probe six-fold fermions in PdSb$_2$, which correspond to the $R$-$X$-$M$ high-symmetry measurement plane in the BZ ($k_z = \pi$), as shown in Figs. 3(a-b). Qualitatively, our experimental data are in a good agreement with our first-principles calculations in both high-symmetry directions, highlighting that the observed nodal point at $R$ indeed consists of a six-fold crossing of three doubly degenerate bands. Moreover, this node is found to be located at a binding energy of ∼ 0.24 eV, which is close to the calculation (0.26 eV). To gain further insight into the band structure around $R$, we demonstrate in Fig. 3(c) the photoemission intensity plot taken with a $h\upsilon$ = 46 $eV$ photon, which is expected to directly probe the $R$-$X$-$M$ high-symmetry plane as well. Owning to the higher momentum/energy resolution under relatively low-energy photons, one can resolve two electron-like bands [labeled as 1 and (2, 3)] centered at the $R$ point at a binding energy $E_b$ ≈ 0.24 $eV$, which is in remarkable accordance with the predicted six-fold fermion band dispersion. Note that it is rather challenging to resolve bands 2 and 3 experimentally, as they only differ by ∼ 2 meV in energy according to the calculation. Some shoulder-like bands (located at - 0.2 ∼ - 0.3 $eV$ below $E_F$) observed in the Fig. 3(c) are probably induced by the folded spectral weight from the $M$ point, which might be due to the $k_z$-broadening effect in the low-energy photon photoemission process. Fig. 3(d) shows the photoemission intensity plot taken with $h\upsilon$ = 33 $eV$ photons, which correspond to the $k_z$ = 0 plane in the BZ. With this photon energy, one can more clearly recognize that the shoulder-like dispersion observed in Fig.3 (c). Intriguingly, here we can still distinguish the characteristic six-fold crossing of three doubly degenerate bands, in which the predicted splitting between band 1 and degenerate bands 2 and 3 could be well resolved. This finding further confirms the $k_z$-broadening effect under low-energy photons and provides more compelling evidence of the existence of six-fold exotic fermions in PdSb$_2$.

To investigate energy dispersion of bands near an six-fold degeneracy, we next demonstrate stacked plots of constant energy contours taken at different binding energies in Fig. 3(e), which clearly show the parabolic dispersion of bands forming the six-fold crossing. Note that this node is distinctly different from that of previously reported spin-1 six-fold fermions, bands involved in which disperse almost linearly towards $E_F$ [21–23]. Figs. 3(f), 3(g) and 3(h) show experimental constant-energy surfaces taken at Fermi level, -100 $meV$ and -240 $meV$, respectively. Their resulting four-petal flower-shape surfaces centered around $R$ are in excellent agreement with the bulk band calculations, further confirming the position and band dispersion of six-fold fermion at $R$. Here, experimental constant-energy surfaces also show some extra features other than those formed by the six-fold fermions, and we would attribute that to spectral projection from adjacent surfaces owing to $k_z$-broadening effects. Overall, the good agreement between our experimental and calculated band structure unambiguously confirms the existence of six-fold fermions in PdSb$_2$.

Different from chiral fermion materials, PdSb$_2$ preserves $D_3d$ symmetry in the little group of the $R$ point, which renders its effective Hamiltonian to be quadratic when close to its six-fold degeneracies [14]. Consequently, PdSb$_2$ should have only bands of zero net Berry flux, and the resulting six-fold fermions should be with C = ± 0, in sharp contrast to double spin-1 fermions with C = ± 2. To further investigate the topological nature of PdSb$_2$, we compare the experiment bulk and surface electronic structure of the (001) cleavage plane with slab calculation around the Fermi level. The experimental Fermi surface is in excellent agreement with the calculation as illustrated in Fig. 4(a) (see more details in Fig. S4). Unprotected surface states can be well separated from bulk states along with the raising of constant energy according to our calculation [Fig. 4(b), (c)], which are similar to fragile surface states discovered in PtBi$_2$ [with the same space group (205) as PdSb$_2$] [35]. However, no topologically protected Fermi arcs connecting degenerate nodes as in chiral fermiions can be observed in PdSb$_2$ in the vicinity of $E_F$, further confirming that there is no abelian Berry curvature (Chern number) associated with these degeneracies.

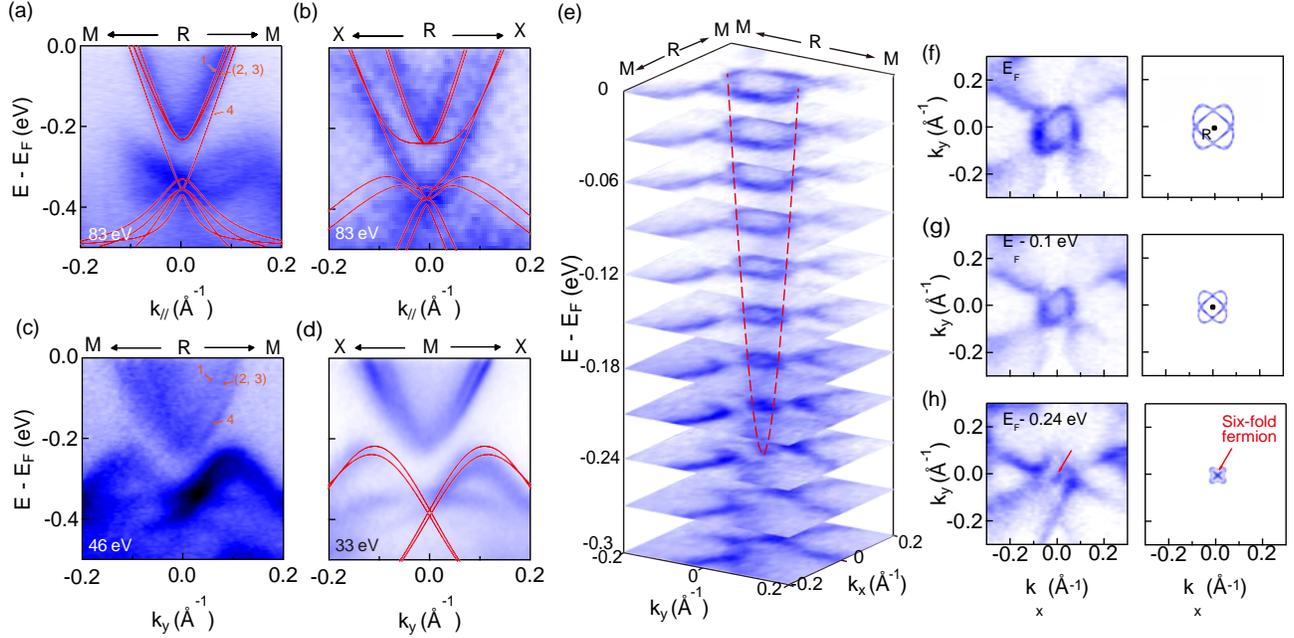

FIG. 3. Bulk electronic structure of PdSb$_2$. (a) Experimental and calculated band dispersion along $M - R - M$ high symmetry direction, measured with $h\nu = 83$ $eV$. (b) Experimental and calculated band dispersion along $X - R - X$ high symmetry direction, measured with $h\nu = 83$ $eV$. For comparison, we plot the calculated bands along $M - R - M$ and $X - R - X$ as red lines on top of the experimental data (a) and (b). Note that the band crossing exhibits three degenerate bands (labeled 1,2,3), which are doubled by the presence of time-reversal symmetry reasulting in six-fold fermion. (c) Experimental band dispersion along $M - R - M$ high symmetry direction, measured with $h\nu = 46$ $eV$. The bands are labeled 1, 2, 3 and 4, corresponding to the calculated result in (a). (d) Experimental and calculated band dispersion along $X - M - X$ high symmetry direction, measured with $h\nu = 33$ $eV$. (e) Energy evolution of the constant-energy surfaces at the R point, taken with $h\nu = 83$ $eV$. Red dashed lines are a guide for the eye. (f) Comparison between the experimental and bulk calculated Fermi surface, measured with $h\nu = 83$ $eV$. (g), (h) Comparison between the experimental and bulk calculated constant-energy surface, which are 0.1 and 0.24 $eV$ below $E_F$, respectively.

In conclusion, by using ARPES measurements combined with first-principles calculations, we verify the existence of six-fold fermions in pyrite-type PdSb$_2$, which are stabilized by its nonsymmorphic symmetry. Through systematic characterization and analysis of the band structures of the bulk and surface states of PdSb$_2$, we have confirmed that the nodes at $R$ in PdSb$_2$ have no chirality (C = 0), which is also guaranteed by the inversion symmetry of its space group. This is the first direct spectroscopic evidence of the presence of six-fold fermions in PdSb$_2$. Our work complements a puzzle of the topological family and provides a good platform for further studies into exotic fermions.

We acknowledge T. Xu and W. W. Kong for fruitful discussion. This work was supported by the National Key R&D Program of the MOST of China (Grant No. 2016YFA0300204) and the National Science Foundation of China (Grant Nos. 11574337, 11874199, 11874263 and 11874264). P. W. was supported by the National Basic Research Program of China (Grant No. 2015CB654901). Y. F. Guo acknowledge the support by the Natural Science Foundation of Shanghai (Grant No. 17ZR1443300), the starting grant of ShanghaiTech University and the Program for Professor of Special Appointment (Shanghai Eastern Scholar). Part of this research used Beamline 03U of the Shanghai Synchron Radiation Facility, which is supported by ME2 project under contract No. 11227902 from National Natural Science Foundation of China. D.W.S. is also supported by "Award for Outstanding Member in Youth Innovation Promotion Association CAS".

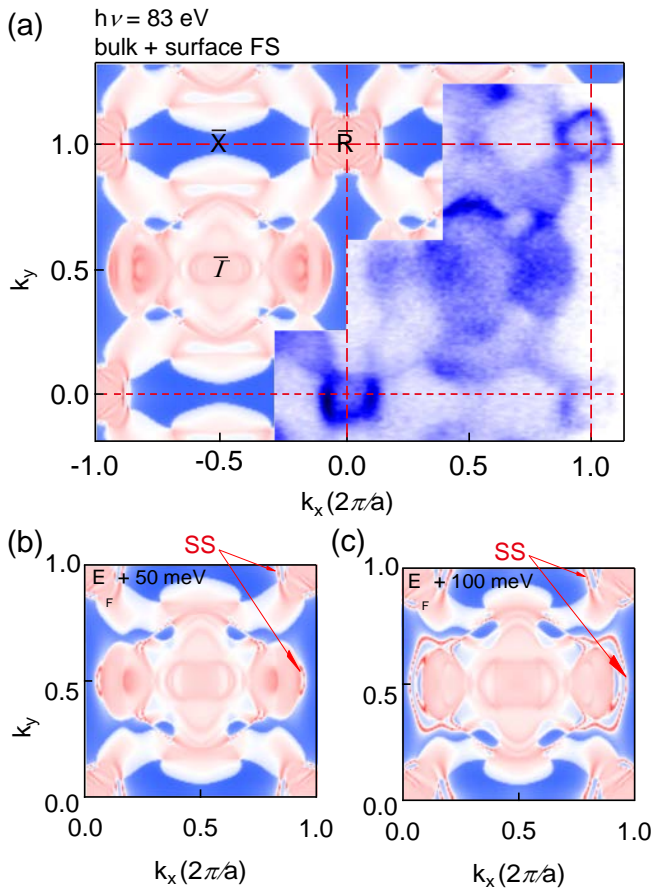

FIG. 4. Bulk electronic structure and surface state spectrum on the (1) cleavage plane. (a) Experimental Fermi surface map measured with $h\nu = 83\ eV$, integrated over 10 $meV$ at the Fermi level. Also shown is calculated Fermi surface spectrum, with high symmetry points marked. (b) and (c) present the calculated surface state spectrum in 2D BZ, which are 50 $meV$ and 100 $meV$ above Fermi surface, respectively. The red arrows in (b) and (c) point to the surface states (SS).